\documentstyle[epsf,prl,aps,twocolumn,floats,epsfig]{revtex}
\begin{document} 
\title{Measuring Information Transfer}
\author{Thomas Schreiber\\
   {\em Max Planck Institute for Physics of Complex Systems, 
   N\"othnitzer Str.~38, 01187 Dresden, Germany}}
\date{\today}
\maketitle
\begin{abstract} 
An information theoretic measure is derived that quantifies the statistical
coherence between systems evolving in time. The standard time delayed mutual
information fails to distinguish information that is actually exchanged from
shared information due to common history and input signals. In our new
approach, these influences are excluded by appropriate conditioning of
transition probabilities. The resulting {\em transfer entropy} is 
able to distinguish driving and responding elements and to
detect asymmetry in the coupling of subsystems.  
\end{abstract}

The time evolution of a system may be called irregular if it generates
information at a non-zero rate. For stochastic or deterministically chaotic
systems, this is quantified by the entropy. For a system consisting of more
than one component, important information on its structure can be obtained by
measuring to which extent the individual components contribute to information
production and at what rate they exchange information among each other. This
paper proposes a method to answer the latter question on the basis of time
series observations.

Many authors have used  {\em mutual information}~\cite{shannon} to quantify the
overlap of the information content of two (sub-) systems. Unfortunately, mutual
information neither contains dynamical nor directional information. Introducing
a time delay in one of the observations is an important, if somewhat arbitrary,
improvement  in  this  respect,  but  still  does  not  explicitly  distinguish
information that is actually exchanged from that due to the
response to a common input signal or history.

The purpose of this paper is to motivate and derive an alternative information
theoretic measure, to be called {\em transfer entropy}, that shares some of the
desired properties of mutual information but takes the dynamics of information
transport into account. With minimal assumptions about the dynamics of the
system and the nature of their coupling one will be able to quantify the
exchange of information between two systems, separately for both directions,
and, if desired, conditional to common input signals.

This work may be seen in the context of a considerable number of  
recently proposed measures~\cite{adhoc} for the nonlinear coherence of signals,
used to study generalized synchronization phenomena in many contects,
most notably in physiological systems. While these measures are often very
powerful for a specific set of applications, it is also important to aim at an
understanding of the underlying theoretical concepts.  In the generic
case that neither of the systems, nor their coupling may be assumed to be
deterministic, information theory seems to be an appropriate
starting point.

Let us briefly recall the most basic concepts of information theory. The
average number of bits needed to optimally encode independent draws of the
discrete variable $I$ following a probability distribution $p(i)$ is given by
the Shannon entropy~\cite{shannon}
\[
    H_I=-\sum_i\, p(i) \log_2 p(i)
\]
where the sum extends over all states~$i$ the process can assume. The base of
the logarithm only determines the units used for measuring information and 
will be dropped henceforth.

In order to construct an optimal encoding that uses just as many bits as given
by the entropy, it is necessary to know the probability distribution~$p(i)$.
The excess number of bits that will be coded if a different distribution $q(i)$
is used instead of~$p(i)$ is given by the Kullback entropy~\cite{Kull}
$K_I=\sum_i\, p(i) \,\log\, p(i)/ q(i)$.
We will later also need the Kullback entropy for conditional probabilities
$p(i|j)$. For a single state $j$ we have $K_j=\sum_i\, p(i|j) \,\log\, 
p(i|j)/ q(i|j)$. Summation over $j$ with respect to $p(j)$ yields
\begin{equation}\label{eq:kull2}
   K_{I|J}=\sum_{i,j}\, p(i,j) \,\log\, {p(i|j) \over q(i|j)}
\,.\end{equation}

The {\em mutual information} of two processes $I$ and $J$ with joint
probability $p_{IJ}(i,j)$ can be seen as the excess amount of code produced by
erroneously assuming that the two systems are independent, i.e. assuming
$q_{IJ}(i,j)=p_I(i)\,p_J(j)$ instead of $p_{IJ}(i,j)$. 
The corresponding Kullback entropy is
\begin{equation}\label{eq:mutual}
   M_{IJ}=\sum\, p(i,j) \,\log\, {p(i,j) \over p(i)\;p(j)}
\end{equation}
which is the well known formula for the mutual information.  Here and in the
following, we omitted the summation index and the subscript of the
probabilities specifying the process. This derivation shows that mutual
information is a natural way to quantify the deviation from independence
of two processes. We have $M_{IJ}=H_I+H_J-H_{IJ}\ge 0$. Note that $M_{IJ}$ is
symmetric under the exchange of $I$ and $J$ and therefore does not contain any
directional sense.

A related, non-symmetric quantity is the conditional entropy $H_{I|J}=-\sum\,
p(i,j)\, \log\, p(i|j) = H_{IJ}-H_J$. However, since $H_{I|J}-H_{J|I}=H_I-H_J$,
it is non-symmetric only due to the different individual entropies and not due
to information flow. Mutual information can be given a directional sense in a
somwhat ad-hoc way by introducing a time lag in either one of the variables and
compute e.g.  
\[
   M_{IJ}(\tau)=\sum\, p(i_n,j_{n-\tau}) \,\log\, 
      {p(i_n,j_{n-\tau}) \over p(i)\;p(j)}
\,.\] 
As we will see below, considering the two systems at different
times occurs naturally as soon as transition probabilities are introduced.
This will yield a more justified approach to measuring information transfer
that explicitly incorporates directional, dynamical structure.

One can incorporate dynamical structure by studying transition
probabilities rather than static probabilities. Consider a system that may be
approximated by a stationary Markov process of order $k$, that is, the
conditional probability to find $I$ in state $i_{n+1}$ at time $n+1$ observes
$p(i_{n+1}|i_n,\ldots,i_{n-k+1})=p(i_{n+1}|i_n,\ldots,i_{n-k})$.  Henceforth we
will use the shorthand notation $i^{(k)}_n=(i_n,\ldots,i_{n-k+1})$ for words of
length $k$, or $k$ dimensional delay embedding vectors.

The average number of bits needed to encode one additional state of the 
system if all previous states are known is given by the {\em entropy rate} 
\begin{equation}\label{eq:rate}
    h_I = -\sum\,   p(i_{n+1},i^{(k)}_n) \, \log\, p(i_{n+1}|i^{(k)}_n)
\,.\end{equation}
Since $p(i_{n+1}|i^{(k)}_n)=p(i^{(k+1)}_{n+1})/p(i^{(k)}_n)$, this is just the
difference between the Shannon entropies of the processes given by $k+1$ and
$k$ dimensional delay vectors constructed from~$I$: $h_I = H_{I^{(k+1)}} -
H_{I^{(k)}}$. 

If $I$ is obtained by coarse graining a continuous system $X$ at resolution
$\epsilon$, the entropy $H_X(\epsilon)$ and entropy rate $h_X(\epsilon)$ will
depend on the partitioning and in general diverge like $\log\epsilon$ when
$\epsilon\to 0$.
However, for the special case of a deterministic dynamical
system, $\lim_{\epsilon\to 0} h_X(\epsilon)=h_{KS}$ may exist and is then
called the Kolmogorov--Sinai entropy. (For non-Markov systems, also the limit
$k\to\infty$ needs to be taken.) Confusingly, the opposite is true for the
mutual information. For generic noisy interdependence, $\lim_{\epsilon\to 0}
M_{XY}(\epsilon)$ is finite and independent of the partition, but for
deterministically coupled processes, $M_{XY}(\epsilon)$ will diverge as
$\epsilon\to 0$.

The Shannon entropy and its generalization, the mutual information, are
properties of the static probability distributions while the dynamics of the
processes is contained in the transition probabilities. For the study of 
the dynamics of shared information between processes it is therefore
desirable to generalize the entropy rate, rather than Shannon entropy, 
to more than one system. In the next section I will propose such
generalizations, in particular one that is non-symmetric under the exchange of
the two processes.

The most straightforward way to construct a mutual information rate by
generalizing $h_I$ to two processes $(I,J)$ is again by measuring the deviation
from independence. The corresponding Kullback entropy
is sometimes called {\em transinformation} and is still symmetric under the
exchange of $I$ and~$J$. It is therefore preferable to measuring the deviation
from the generalized Markov property 
\[
   p(i_{n+1}|i^{(k)}_n)=p(i_{n+1}|i^{(k)}_n,j^{(l)}_n)
\,.\]
In the absence of information flow from $J$ to $I$, the state of $J$ has no
influence on the transition probabilities on system~$I$. The incorrectness of
this assumption can again be quantified by a Kullback entropy (\ref{eq:kull2})
by which we define the {\em transfer entropy}:
\begin{equation}\label{eq:transfer}
   T_{J\to I} = \sum\,
         p(i_{n+1},i^{(k)}_n,j^{(l)}_n)\, \log \,
         {p(i_{n+1}|i^{(k)}_n,j^{(l)}_n)\over p(i_{n+1}|i^{(k)}_n)}
\,.\end{equation}
This is the central concept of this paper. The most natural choices for $l$ are
$l=k$ or $l=1$. Usually, the latter is preferable for computational
reasons. $T_{J\to I}$ is now explicitly non-symmetric since it measures the
degree of dependence of $I$ on $J$ and not vice versa.

For coarse grained states $(I,J)$ of continuous systems $(X,Y)$, the limit
$\lim_{\epsilon\to 0} T_{Y\to X}(\epsilon)$ is finite and independent of the
partition, except for the case of deterministic coupling, when $T_{Y\to
X}(\epsilon)$ diverges as $\epsilon\to 0$. In this respect, transfer entropy
behaves like mutual information.  If computationally feasible, the influence of
a known common driving force $Z$ may be excluded by conditioning the
probabilities under the logarithm to $z_n$ as well.

For numerical and practical applications, the limit $\epsilon\to 0$ is not
obtainable and has to be replaced appropriately. Either one can study transfer
entropy as a function of the resolution, or one can fix a resolution for the
scope of a study. Furthermore, there are several methods of coarse graining and
a partition consisting of a fixed mesh of boxes is not always the best choice.
Fixed boxes are only suitable in cases where data can be produced with little
effort and small statistical errors at reasonable speed of computation are
desired.

For time series applications, an alternative implementation
using generalized correlation integrals is preferable. Mutual information and
redundancies have been generalized for their estimation by order $q$
correlation integrals~\cite{generalized}. It is possible to follow the
same arguments in generalizing transfer entropy. However, for the
computationally most attractive case $q=2$, we would have to give up positivity
of~$T_{I\to J}$. Instead, we propose an implementation 
of the definition~(\ref{eq:transfer}) where the probability 
measure $p(i_{n+1},i^{(k)}_n,j^{(l)}_n)$ is realized by a sum over all
available realizations of $(x_{n+1},x^{(k)}_n,y^{(l)}_n)$ in a time series.
The transition probabilities are expressed by joint probabilities and then
obtained by kernel estimation, e.g.
\[
   \hat{p}_r(x_{n+1},x_n,y_n)=
      {1\over N} \sum_{n'}
        \Theta\left(\; 
                 \left|
            \left(\!\!\begin{array}{c}x_{n+1}-x_{n'+1}\\x_n-x_{n'}\\y_n-y_{n'}
                      \end{array}\!\!\right)
                 \right|
                    -r
              \right) 
.\]
We use the step kernel $\Theta(x>0)=1$; $\Theta(x\le 0)=0$.  The norm $|\cdot|$
can be simply the maximum distance but other norms and kernels can be
considered. In particular, different overall scales of $X$ and $Y$ can be
accounted for by using appropriate weights.  Similarly to standard dimension
and entropy calculations, fast neighbour search strategies are advisable for
all but the smallest data sets. Dynamically correlated pairs should be excluded
as usual. Since these technical issues are the same as in many nonlinear time
series methods, the reader is referred to the discussion in the
literature~\cite{ourbook}.

\begin{figure}
\centerline{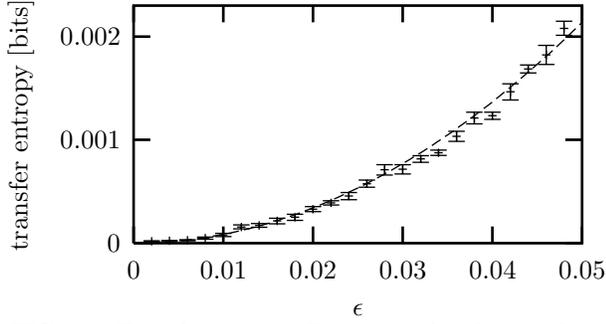}
\caption[]{\label{fig:tent}
   Transfer entropy $T_{I^{m-1} \to I^m}$ for the coupling direction as a
   function of the coupling strength $\epsilon$ in a tent map lattice (binary
   partition).  Errorbars: error of the mean of 10 runs of 100000
   iterates. Line: theoretical curve $\alpha^2\epsilon^2/\ln(2)$ with fitted
   $\alpha=0.77$.  }
\end{figure}

In order to demonstrate the use of transfer entropy, let us study three
examples, two spatio-temporal systems and a bi-variate physiological time
series. In a one dimensional lattice of unidirectionally coupled maps
\begin{equation}
\label{eq:cml}
    x_{n+1}^m = f( \epsilon x_n^{m-1} + (1-\epsilon) x_n^m)
\,,
\end{equation}
information can be transported only in the direction of increasing~$m$. One of
the simplest cases is given by the tent map, $f(x<0.5)=2x$; $f(x\ge
0.5)=2-2x$. Let us study coarse grained states $I^m$ with $i_n^m$ defined by a
partition at $x_0=0.5$. At zero coupling, all static and transfer probabilities
are equal to 1/2, $M(\tau) = 0$ for all values of $\tau$, and also $T_{I^{m-1}
\to I^m} = T_{I^{m} \to I^{m-1}} = 0$.  For nonzero coupling, we still have
$T_{I^{m} \to I^{m-1}} = 0$, but $T_{I^{m-1} \to I^m}$ becomes positive.  For
small coupling, it can be assumed that the invariant density at a single site
is essentially unchanged whence the transition probabilities
$p(I_{n+1}^m|I_{n}^m,I_{n}^{m-1})$ are changed by an amount proportional
to~$\epsilon$. In particular, $p(0|0,0)$, $p(0|1,1)$, $p(1|0,1)$, and
$p(1|1,0)$ are increased by a factor $1+\alpha\epsilon$ with $\alpha=O(1)$. All
others are decreased by that amount.  Evaluating (\ref{eq:transfer}) in lowest
order of $\epsilon$ with $k=l=1$, we obtain $T_{I^{m-1} \to
I^m}=\alpha^2\epsilon^2/\ln(2)+O(\epsilon^4)$.  For this particular case, the
changes in $p(i_{n+1}^m,i_n^{m-1})$ exactly cancel out and the mutual
information is zero.  Figure~\ref{fig:tent} shows a numerical verification of
these results for a spatially periodic lattice of 100 maps. Averages of 10 runs
of $10^5$ iterates after $10^5$ transients are shown. The transfer entropy
$T_{I^{m} \to I^{m-1}}$ and both directions of $M(\tau=1)$ were found
consistent with zero and are therefore not shown.

\begin{figure}
\centerline{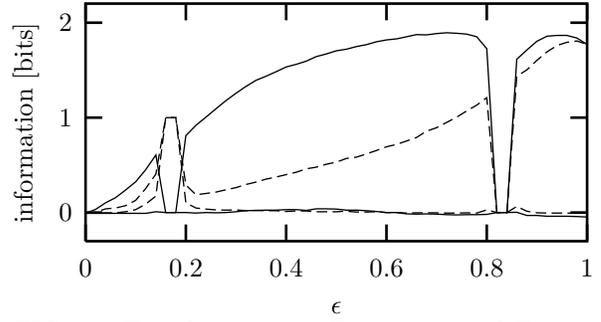}
\caption[]{\label{fig:one}
   Transfer entropies $T_{X^{m-1} \to X^m}$ and $T_{X^{m} \to X^{m-1}}$ (solid
   lines) and time delayed mutual information $M_{X^1,X^2}(\tau=1)$ and
   $M_{X^2,X^1}(\tau=1)$ (dashed lines) as functions of the coupling strength
   $\epsilon$ for a unidirectionally coupled Ulam lattice. For both quantities,
   the upper line denotes the direction $X^{m-1}\to X^m$ while the
   lower line shows $X^{m+1}\to X^m$. Although the lattice undergoes a
   sequence of bifurcations, the transfer entropy $T$ clearly reflects the
   unidirectional character of the coupling. It also consistently outperforms
   the time delayed mutual information in this respect. See text for further
   details.}
\end{figure}

The situation is more complicated for the Ulam map $f(x)=2-x^2$ and non-small
coupling.  For each coupling, a bi-variate time series was generated using a
lattice of $100$ points (random initial conditions) and recording 10000
iterates of $x_n^1$ and $x_n^2$ after $10^5$ steps of transients.  Correlation
sums at $r=0.2$ were used to compute mutual information in both directions,
$M_{X^1,X^2}(\tau=1)$ and $M_{X^2,X^1}(\tau=1)$, as well as transfer entropies
$T_{X^1\to X^2}$ and $T_{X^2\to X^1}$ with $k=l=1$. Neighbors closer in time
than 100 iterates were excluded from the kernel estimation.

Figure~\ref{fig:one} shows $M$ and $T$ as functions of the coupling strength. 
Both $M$ and $T$ are able to detect the anisotropy since the
information is consistently larger in the positive direction. The lattice
undergoes a number of bifurcations when the coupling is changed.  Around
$\epsilon=0.18$, the asymptotic state is of temporal and spatial period
two. For this case, the mutual information is found to be 1~bit. This is
correct although information is neither produced nor exchanged and reflects the
static correlation between the sites. The transfer entropy finds a zero rate of
information transport, as desired. Around this pariodic window, the mutual
information is non-zero in both directions and the signature of the
unidirectional coupling is less pronounced. Around $\epsilon=0.82$, the
lattice settles to a (spatially inhomogenious) fixed point state. Here both
measures correctly show zero information transfer. The most important finding,
however, is that the transfer entropy for the negative direction remains
consistent with zero for all couplings, reflecting the causality in the system.

\begin{figure}[t]
\centerline{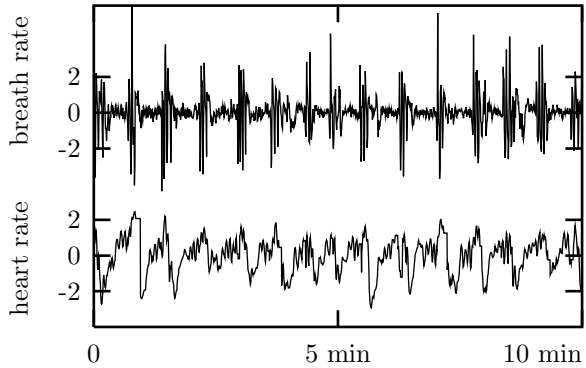}
\caption[]{\label{fig:bdat_series}
   Bi-variate time series of the breath rate (upper) and instantaneous heart
   rate (lower) of a sleeping human. The data is sampled at 2~Hz. Both traces
   have been normalized to zero mean and unit variance.}
\end{figure}

\begin{figure}
\centerline{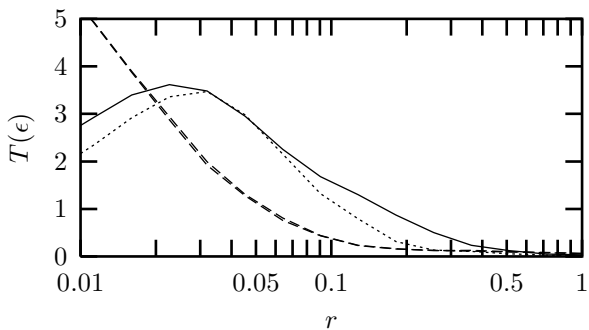}
\caption[]{\label{fig:bdat}
   Transfer entropies $T(\makebox{heart $\to$ breath})$ (solid line),
   $T(\makebox{breath $\to$ heart})$ (dotted line), and time delayed mutual
   information $M(\tau=0.5~{\rm s})$ (directions indistinguishable, dashed
   lines) for the physiological time series shown in
   Fig.~\ref{fig:bdat_series}. }
\end{figure}

As a last example, take a bi-variate time series (see
Fig.~\ref{fig:bdat_series}) of the breath rate and instantaneous heart rate of
a sleeping human suffering from sleep apnea (part of data set B of the Santa Fe
Institute time series contest held in
1991~\cite{Rigney}). Figure~\ref{fig:bdat} shows that while time delayed mutual
information is almost symmetric between both series, the transfer entropy
indicates a stronger flow of information from the heart rate to the breath rate
than vice versa over a significant range of length scales~$r$.  Note that for
small $r$, the curves deflect down to zero due to the finite sample size. This
result is consistent with the observation that the patient breathes in bursts
which seem to occur whenever the heart rate crosses some threshold. Certainly,
both signals could instead be responding to a common external trigger.

In conclusion, the new {\em transfer entropy} is able to detect the directed
exchange of information between two systems. Unlike mutual information, it is
designed to ignore static correlations due to the common history or common
input signals. Most prominent applications include multivariate analysis of
time series and the study of spatially extended systems.

Several authors~\cite{swinney} have proposed to use time delayed mutual
information $M(\Delta l,\tau)$ as a function of spatial distance $\Delta l$ and
temporal delay $\tau$ to define a velocity of information transport in
spatio-temporal systems. Often, one finds that $M(\Delta l,\tau)$ for fixed
$\Delta i$ reaches a local maximum at some lag $\tau^*$. Hence a velocity can
be defined by the ratio $\Delta i/\tau^*$, in particular if that ratio is
fairly constant over the resolvable range of values for~$\Delta i$. This
reasoning has been challenged~\cite{diplom} by giving an example where the
above interpretation implies super-luminar communication. In fact, much of the
common information is due to the common history that allows the lattice to
partially synchronize.  Preliminary results indicate that appropriate
conditioning for the common history by replacing time delayed mutual
information by a variant of Eq.(\ref{eq:transfer}) resolves this aparent
paradox.  However, conditioning with respect to a large number of variables
poses immense numerical problems whence this study will be concluded at a later
time.

Part of this work has been supported by the SFB 237 of the Deutsche
Forschungsgemeinschaft.

\end{document}